%% file: main_text.tex
\documentclass{Nature}
\usepackage{hyperref}
\usepackage{graphicx}
\usepackage{epsfig}
\usepackage{pdfpages}
\usepackage{times}
\usepackage{todonotes}
\usepackage{authblk}
\usepackage[hyphenbreaks]{breakurl}
\usepackage{amssymb}
\usepackage{amsthm}
\usepackage{xcolor}
\usepackage[normalem]{ulem}
\usepackage[labelformat=empty]{caption}
\usepackage{algorithm2e}
\usepackage{algorithmic}
\usepackage{subfigure}

\SetKwInput{Continue}{continue}
\input{math_commands}

\newcommand{\model}[0]{DIRAC}
\newcommand{\modelMetaQ}[0]{DIRAC$^1$}
\newcommand{\modelMultiQ}[0]{DIRAC$^m$}
\newcommand{\modelPT}[0]{PT$_{\rm DIRAC}$}
\newcommand{\ENCmodel}[0]{SGNN}

\newcommand{\nop}[1]{}
\title{Finding spin glass ground states through  deep reinforcement learning}
\author{Changjun Fan$^{1*}$, Mutian Shen$^{2*}$, Zohar Nussinov$^{2}$, Zhong Liu$^{1}$, Yizhou Sun$^{3}$, Yang-Yu Liu$^{4}$}

\begin{document}
\maketitle

 \begin{affiliations}
      \item College of Systems Engineering, National University of Defense Technology, Changsha, 410073, China
      \item Department of Physics, Washington University in St. Louis, Campus Box 1105, 1 Brookings Drive, St. Louis, MO 63130, USA
      \item Department of Computer Science, University of California, Los Angeles, CA, 90024, USA
      \item Channing Division of Network Medicine, Department of Medicine, Brigham and Women's Hospital and Harvard Medical School, Boston, MA, 02115, USA
      \\ \noindent *These authors contributed equally to this work.
 \end{affiliations}
\vspace{0.5in}

\begin{abstract}
\input{main_text_section/abstract}
\end{abstract}

\input{main_text_section/introduction}
\input{main_text_section/model}
\input{main_text_section/results}
\input{main_text_section/conclusion}

\bibliographystyle{Nature}
\bibliography{ref}

\clearpage
\begin{addendum}

\item[Author contributions] 
Y.-Y.L. conceived the project. Y.-Y.L. and Y.S. designed and managed the project. F.C. and M.S. contributed equally to this work. F.C. and M.S. performed all the numerical calculations and analyzed the results, Y.-Y.L., Y.S., Z.N., and Z.L. interpreted and improved the results. F.C., M.S., and Y.-Y.L. wrote the manuscript, Y.S., Z.N., and Z.L. edited the manuscript.

\item[Competing interests] The authors declare that they have no competing interests.
 
\item[Correspondence] Correspondence and requests for materials should be addressed to Y.-Y.L. (yyl@channing.harvard.edu) and Y.S. (yzsun@cs.ucla.edu).
\end{addendum}

\input{main_text_section/figures}
\end{document}

%% file: math_commands.tex
\usepackage{amsmath,amsfonts,bm}

\def\eqref#1{equation~\ref{#1}}
\def\1{\bm{1}}

\newcommand{\zb}{\mathbf{z}}

\newcommand{\cB}{\mathcal{B}}

\DeclareMathAlphabet{\mathsfit}{\encodingdefault}{\sfdefault}{m}{sl}
\SetMathAlphabet{\mathsfit}{bold}{\encodingdefault}{\sfdefault}{bx}{n}

%% file: main_text_section/abstract.tex
%meaning and challenge
Spin glasses are disordered magnets with random interactions that are, generally, in conflict with each other. Finding the ground states of spin glasses is not only essential for the understanding of the nature of disordered magnetic and other physical systems, but also useful to solve a broad array of hard combinatorial optimization problems across multiple disciplines. Despite decades-long efforts, an algorithm with both high accuracy and high efficiency is still lacking. Here we introduce \model~ --- a deep reinforcement learning framework, which can be trained purely on small-scale spin glass instances and then applied to arbitrarily large ones. \model~displays better scalability than other methods and can be leveraged to enhance any thermal annealing method. Extensive calculations on 2D, 3D and 4D Edwards-Anderson spin glass instances demonstrate the superior performance of \model~over existing methods. As many hard combinatorial optimization problems have Ising spin glass formulations, our results suggest a promising tool in solving these hard problems. Moreover, the presented algorithm will help us better understand the nature of the low-temperature spin-glass phase, which is a fundamental challenge in statistical physics. 

%% file: main_text_section/introduction.tex
\section*{Introduction}
The Ising spin glass is a classical disordered system that has been studied for decades~\cite{binder1986spin,mezard1987spin}. Its spectacular behaviors have attracted considerable interests in several branches of science, including physics, mathematics, computer science, and biology. The endogenous nature of quenched disorder in spin glasses results in the fact that, it is hard to find out the ground state of such a system due to the frustrations (i.e., the impossibility of simultaneously minimizing all the interactions), despite its seemingly simple Hamiltonian~\cite{hartmann2002optimization}: 
\begin{equation}
    \mathcal{H} = - \sum_{\langle i, j \rangle} J_{ij} \sigma_i \sigma_j.
\end{equation}
In general, this Hamiltonian can be defined on arbitrary graphs. Here, we will focus on the most heavily studied lattice realization of the nearest neighbor Ising spin glass in which the sites lie on a $d$-dimensional hypercubic lattice with $N=L^d$ sites (see Fig.~\ref{fig:case_study} for 2D instances) and $\sigma_i=\pm 1$ represents the binary Ising spin value at site $i$. The coupling $J_{ij}$ is a Gaussian random variable that represents the interaction strength between two neighboring spins $i$ and $j$. In the literature, this is often referred to as the Edwards-Anderson (EA) spin glass model. The EA model aims at capturing the quintessential character of real, physically occurring, spin glasses~\cite{sherrington1975solvable}. Comparing to other short range models such as the mean-field Bethe Lattice~\cite{Chayes86amean}, the EA model seems more challenging in the sense that there exists amounts of short loops that will lead to much more frustration.

There are at least three strong motivations to find the ground states of spin glasses. First of all, finding the spin glass ground states is a key to the mysteries behind the strange and complex behaviors of spin glasses (and many other disordered systems), such as its glassy phase~\cite{ceccarelli2011ferromagnetic} and ergodicity breaking~\cite{cugliandolo1995weak}. In particular, ground-state energies in different boundary conditions can be used to compute the stiffness exponent of spin glasses, which can help us ensure the existence of a spin glass phase at finite temperature~\cite{carter2002aspect, hartmann1999scaling}. Second, finding ground states of Ising spin glasses in three or higher dimensions is a non-deterministic polynomial-time (NP) hard problem~\cite{barahona1982computational}, which is closely related to many other hard combinatorial optimization problems~\cite{lucas2014ising}. For example, all  of Karp's 21 NP-complete problems and many NP-hard problems (such as the max-cut problem, the traveling salesman problem, the protein folding problem, etc.) have Ising spin glass formulations~\cite{lucas2014ising}. Therefore, finding the Ising spin glass ground states may help us solve many other NP problems. Finally, the celebrated Hopfield model~\cite{Hopfield} and other pioneering models of neural networks drew deep connections with Ising magnets~\cite{Little} (and spin glasses, in particular~\cite{Amit,Sompolinsky}) on general networks. Also, finding the underlying community structure of general networks can be mapped onto a spin glass ground state problem~\cite{Fortunato,RB}. The study of spin glasses and their ground states has led to (and will continue lead to) the development of powerful optimization tools such as the cavity method and Belief Propagation that will further shed new light on computational complexity transitions~\cite{mezard1987spin,BP1}. 

% challenges
Given the NP-hard nature of finding the spin glass ground states in three or higher dimensions, the exact branch-and-bound approach can only be used for very small systems~\cite{de1995exact}. For two-dimensional lattices with periodic boundary conditions in at most one direction (or planar graphs in general), the Ising spin glass ground states can be calculated by mapping to the minimum-weight perfect matching problem, which can be exactly solved in polynomial time~\cite{hartmann2011ground, weigel2018}. However, for general cases with large system sizes, we lack a method with both high accuracy and high efficiency. We used to rely on heuristic methods. In particular, heuristic Monte Carlo methods based on thermal annealing, e.g., simulated annealing~\cite{kirkpatrick1983optimization}, population annealing~\cite{gubernatis2003monte} and parallel tempering~\cite{swendsen1986replica, geyer1991computing, hukushima1996exchange, earl2005parallel}, have been well studied in the statistical physics community.   

Recently, reinforcement learning (RL) has proven to be a promising tool in tackling many combinatorial optimization problems, such as the minimum vertex cover problem~\cite{mnih2015human}, the minimum independent set problem~\cite{li2018combinatorial}, the network dismantling problem~\cite{fan2020finding}, the travelling salesman problem~\cite{bello2016neural}, the vehicle routing problem~\cite{nazari2018reinforcement}, etc. Compared to traditional methods, RL-based algorithms can achieve a more favorable trade-off between accuracy and efficiency in solving those combinatorial optimization problems. On the other hand, despite the fact that AI has been frequently used to 
facilitate physics research~\cite{udrescu2020ai}, recent work also shows that reinforcement learning, specifically, can be applied to solve statistical mechanics, for example, by computing variational free energy~\cite{PhysRevLett.122.080602}. We also note that RL was recently used to devise a smart temperature control scheme of simulated annealing in finding ground states of the 2D spin glass system, which enabled small systems to better escape local minimum and reach their ground states with high probability~\cite{mills2020finding}. However, this RL-enhanced simulated annealing still fails in finding ground states for larger spin glass systems in three or higher dimensions. %Instead, we propose to leverage RL 
In this work, we introduce \model~(\textbf{\underline{D}}eep reinforcement learning for sp\textbf{\underline{I}}n-glass g\textbf{\underline{R}}ound-st\textbf{\underline{A}}te \textbf{\underline{C}}alculation), an RL algorithm that can directly calculate spin glass ground states. \model~has several advantages. First, it displays better scalability than other methods. Second, it demonstrates superior performances (in terms of accuracy) over the state-of-the-art thermal annealing methods, e.g., parallel tempering. Finally, it can be leveraged to enhance any thermal annealing method and offer better solutions.

%% file: main_text_section/model.tex
\section*{Results}
\subsection*{Reinforcement learning formulation}
Following many other RL frameworks in solving combinatorial optimization problems~\cite{silver2016mastering, khalil2017learning, bello2016neural, mazyavkina2021reinforcement}, \model~considers the spin glass ground state search as a Markov decision process (MDP), which involves an agent interacting with the environment, and learning an optimal policy that sequentially takes the long-sighted action so as to accumulate the maximum rewards. To better describe this process, we first define \emph{state}, \emph{action} and \emph{reward} in the context of Ising spin glass ground state calculation. \emph{State}: a state $s$ represents the observed spin glass instance, including both the spin configuration $\{\sigma_i\}$ and the coupling strengths $\{J_{ij}\}$, based on which the optimal action will be chosen. The \emph{terminal state} $s_{\rm T}$ is met when the agent has tried to flip each spin once. 
\emph{Action}: an action $a^{(i)}$ means to flip spin $i$. \emph{Reward}: the reward $r(s,a^{(i)},s')$ is defined as the energy change after flipping spin $i$ from state $s$ to get a new state $s'$: $r(s,a^{(i)},s') = 2\sum_{j \in \partial i} J_{ij} \sigma_i \sigma_j$, where $\partial i$ refers to the nearest neighbors of $i$.

Through the RL formulation, we seek to learn a policy $\pi_{\Theta}(a_t|s_t)$ that takes the observed state $s_t$ and produces the action $a_t$ corresponding to the optimal spin flip at any time step $t$. Here $\Theta=\{\Theta_\mathrm{E}, \Theta_\mathrm{D}\}$ represents a collection of learnable encoding parameters $\Theta_\mathrm{E}$ and decoding parameters $\Theta_\mathrm{D}$, which will be updated through RL.

\subsection*{\model~architecture}
We design \model~to learn the policy $\pi_\Theta$ automatically. As shown in Fig.~\ref{fig:framework}, \model~consists of two phases: offline training and online application. During the offline training phase, \model~is self-taught on randomly generated small-scale EA spin glass instances. For each instance, \model~interacts with the environment through a sequence of \emph{states}, \emph{actions} and \emph{rewards} (Fig.~\ref{fig:framework}~{\bf a}). Meanwhile, \model~gains experiences to update its parameters, which enhances its ability in finding the ground states of EA spin glasses (Fig.~\ref{fig:framework}~{\bf b}-{\bf c}). For online application, the well-trained \model~agent can be used either directly (\modelMetaQ, Fig.~\ref{fig:framework}~{\bf d}) or iteratively (\modelMultiQ, Fig.~\ref{fig:framework}~{\bf e}) or just as a plug-in to any thermal annealing method, on EA spin glass instances with much larger sizes than the training ones.

\model's success is mainly determined by the following two key issues: 1) How to represent \emph{states} and \emph{actions} effectively? 2) How to leverage these representations to compute a $Q$-value, which predicts the long-term gain for an action under a state. We refer to these two questions as the \emph{encoding} and \emph{decoding} problem, respectively.

{\bf Encoding.}
Since a hypercubic lattice can be regarded as a special graph, we design an encoder based on graph neural networks~\cite{kipf2016semi, hanjunnips2017, hamilton2017inductive, velickovic2017graph, gilmer2017neural, xu2018powerful}, namely \ENCmodel~(\textbf{\underline{S}}pin \textbf{\underline{G}}lass \textbf{\underline{N}}eural \textbf{\underline{N}}etwork), to represent states and actions. As shown in Fig.~\ref{fig:SGNN}, to capture the coupling strengths (i.e., edge weights $J_{ij}$), which are crucial to determine the spin glass ground states, \ENCmodel~performs two updates at each layer, specifically, the edge-centric update and the node-centric update, respectively. The edge-centric update (Fig.~\ref{fig:SGNN}~{\bf b}) aggregates edge embedding vectors, which are initialized as edge features, from its adjacent nodes. The node-centric update (Fig.~\ref{fig:SGNN}~{\bf c}) aggregates node embedding vectors, which are initialized as node features, from its adjacent edges. Both updates concatenate the self embedding and the neighborhood embedding and are afterwards subjected to a non-linear transformation (e.g., rectified linear unit, $\mathrm{ReLU}(z)=z$ if $z > 0$ and $0$ otherwise). Traditional graph neural networks architectures often carry one node-centric update~\cite{kipf2016semi, hamilton2017inductive, velickovic2017graph, xu2018powerful}, with edge weights taken as node's neighborhood if needed. Yet this would fail in our case where edge weights play vital roles, and thus lead to unsatisfactory performances.

For a $d$-dimensional hypercubic lattice with periodic boundary conditions, since there are no node attributes and useful node structure statistics (all nodes share the same degree $2d$), it is hard to obtain node features. We here establish a rectangular coordinate system that takes any node in the lattice as the origin, then other nodes' coordinates in this coordinate system are used as their node features. 

\ENCmodel~repeats several layers of both edge-centric and node-centric updates, and finally obtains an embedding vector for each node (or spin) (Fig.~\ref{fig:SGNN}~{\bf e}). Essentially, the node's embedding vector after $K$ layers captures both its position and its long-range couplings with neighbors within $K$ hops (see Fig.~\ref{fig:SGNN} {\bf f} for an example of $K=5$). In our RL setting, each node is subject to a potential action, thus we also call the embedding vector of node $i$, denote as ${\bf z}_i$, as its \emph{action embedding}. Collectively, we denote ${\bf z}_a = \{ {\bf z}_i \}$, which includes embedding vectors for all the nodes $i=1,\cdots,N$. To represent the whole lattice (i.e., the \emph{state} in our setting) and obtain the \emph{state embedding}, denote as ${\bf z}_s$, we sum over all node embedding vectors, which is a straightforward but empirically effective way for graph-level encoding~\cite{hanjunnips2017}.

{\bf Decoding.}
Once the action embedding ${\bf z}_a$ and state embedding ${\bf z}_s$ are computed in the encoding stage, \model~will leverage these representations to define the state-action pair value function $Q(s,a^{(i)}; \Theta)$, which predicts the expected future rewards if taking action $a^{(i)}$ under state $s$, and following the policy $\pi_{\Theta}$ till the end. Specifically, we concatenate the embeddings of state and action, and apply a neural network with nonlinear transformations to map the concatenation to a scalar value. In theory, any neural network architecture can be used. Here for the sake of simplicity, we adopt the classical multilayer perceptron (MLP) with ${\rm ReLU}$:
\begin{equation}\label{Q_score}
    Q(s,a^{(i)}; \Theta) = {\rm MLP} ([\zb_s, \zb_i]; \Theta).
\end{equation}

{\bf Offline training.}
We will adopt the above $Q$ score to design the spin glass ground state search strategy. Prior to that, we first need to optimize the $Q$ function to predict a more accurate future gain, which will be trained offline. Specifically, we minimize the following $n$-step $Q$-learning loss and perform mini-batch gradient parameters updates over large amounts of experience transitions (Fig.~\ref{fig:framework}~{\bf a}):
\begin{equation}\label{loss}
    \mathcal{L} = \mathbb{E}_{(s_t,a_t,r_{t,t+n},s_{t+n})\sim \cB}[(r_{t,t+n}+\gamma \max_{a_{t+n}}Q(s_{t+n},a_{t+n};\hat{\Theta})-Q(s_t,a_t;\Theta))^2],
    % \mathcal{L} = 
\end{equation}
where the 4-tuple transition $(s_t,a_t,r_{t,t+n},s_{t+n})$ is randomly sampled from the experience replay buffer $\cB$,$r_{t,t+n}=\sum_{k=0}^{n-1}r(s_{t+k},a_{t+k},s_{t+k+1})$  represents the $n$-step cumulated reward, $\gamma$ controls how much to discount future rewards. $\hat{\Theta}$ is the target parameter set, which will only be updated with $\Theta$ every a certain number of steps.

{\bf Online application.}
\model~will be trained over a large amount of small random spin glass instances. Once the training phase is finished, we will design the optimized $Q^{\ast}$-based ground state search strategy. Traditional $Q$-learning greedily takes the highest $Q$-value action each step till the end. We refer this policy as the vanilla or original \model~strategy, and denote as \modelMetaQ. As shown in Fig.~\ref{fig:framework}~{\bf d}, similar to the training phase, \modelMetaQ~starts from the all-spin-up configuration $\{\sigma_i = +1\}$ for the input instance, and then greedily flips the highest-$Q$ spin till the end, i.e., the all-spin-down configuration $\{\sigma_i = -1\}$. The spin configuration of the lowest energy encountered during this process is returned as the predicted ground state.
 
Due to the restriction of the finite-horizon MDP, \modelMetaQ~can only start from the uniform spin configuration, e.g., the all-spin-up configuration. To enable \modelMetaQ~to start from arbitrary spin configurations, we employ the so-called \emph{gauge transformation} in such Ising lattice systems~\cite{WegnerGauge}. The gauge transformations between one spin glass instance $\{\sigma_i, J_{ij}\}$ and another instance $\{\sigma'_i, J'_{ij}\}$ are given by~\cite{Ozeki95,PhysRevB.72.045137}
\begin{equation}\label{eq:gauge-transform}
    J'_{ij}= J_{ij}t_it_j, \sigma'_i=\sigma_it_i, 
\end{equation}
where $t_i=\pm 1$ are independent auxiliary variables so that $\sigma'_i$ can take any desired Ising spin value. This naturally leaves the system energy invariant, because $J_{ij}' \sigma_i' \sigma_j' = J_{ij} \sigma_i \sigma_j$. With the aid of gauge transformations shown in Eq.~(\ref{eq:gauge-transform}), we can transform any spin configuration to a uniform configuration. This leads to a much more powerful strategy beyond \modelMetaQ, which we name as \modelMultiQ. As the name shows (also shown in Fig.~\ref{fig:framework}{\bf e}), \modelMultiQ~repeats multiple rounds of \modelMetaQ~until the system energy converges. 

Another prime use of the gauge transformation is the so-called \emph{gauge randomization}~\cite{hamze2018near}, where one may execute many runs (or randomizations) of \model~(either \modelMetaQ~or \modelMultiQ) for an input spin glass instance, with each run the instance is randomly initialized with a different spin configuration. The configuration of the lowest energy among these runs is then returned as the predicted ground state of the input instance. We can see from Fig.~\ref{fig:convergence-comp} that both \modelMultiQ~and gauge randomization greatly improve upon \modelMetaQ.

\emph{\model~plug-in}. \model~can also serve as a plug-in to Monte-Carlo based methods, such as simulated annealing and parallel tempering. The key ingredient of these methods is the so-called \emph{Metropolis-Hasting} criterion:
\begin{equation}
    P(\Delta E;\beta) = \min(1,e^{-\beta \Delta E}),
\end{equation}
which implies that the probability of accepting a move with energy change $\Delta E$ at $\beta$ (indicates the inverse temperature, $1/T$) is the minimum of 1 and $e^{-\beta \Delta E}$. The move is usually referred to as a small perturbation of the system, and in our case, is defined as a single-spin flip. At high temperatures (i.e., lower $\beta$), the Metropolis-Hasting criterion tends to accept all possible moves. However, at low temperatures it is more likely to accept those that could lower the energy (i.e., with $\Delta E <0$), rendering the move-and-accept iteration more like a greedy search. The art of these Monte-Carlo based methods, in some sense, is the balance of explorations at high temperatures and exploitation (energy descents) at low temperatures. The general idea of the plugged-in \model~is to replace the move-and-accept iteration at low temperatures with a single \model~descent, as it has a better performance than an energy descent procedure (Fig.~\ref{fig:energy_gap}).

%% file: main_text_section/results.tex
\subsection*{A long-sighted greediness}
In Fig.~\ref{fig:energy_gap}, we compare the system's energy difference between \modelMetaQ~and the Greedy algorithm at each step. \modelMetaQ~and Greedy both perform in a greedy way. The former greedily flips the highest-$Q$-score spin at each step while the latter greedily flips the greatest-energy-drop spin at each step. Greedy represents an extremely short-sighted strategy, since it focuses only on each step's maximum energy drop. Fig.~\ref{fig:energy_gap} clearly shows that compared to this short-sighted strategy, \modelMetaQ~always goes through a high-energy state temporarily for the early steps, so as to reach a much lower energy state in the long run. In other words, \modelMetaQ~is smart enough to sacrifice in the short term for a long-term gain.

\subsection*{Performance of finding the ground state}
To demonstrate the power of \model~in finding the ground states of Ising spin glasses, we first calculate the probability $P_0$ of finding the ground state of small-scale EA spin glass instances as a function of the number of initial spin configurations $n$ (see Fig.~\ref{fig:gs_compare}). Empirically, we compute $P_0$ as the fraction of 1000 random instances for which the ground state is found at a given $n$. We found that \model~enables a much faster finding of true ground states (which are confirmed by the branch-and-bound based Gurobi solver~\cite{gurobi}) than the Greedy, SA, and the state-of-the-art PT algorithm. In fact, all \model~variants (\modelMultiQ, \modelMetaQ~and \modelPT) facilitate the finding of ground state, with much less initial configurations when compared to the existing algorithms. For example, in the case of $d=2$ and $L=10$ (Fig.~\ref{fig:gs_compare} {\bf a}), denote $n^{\ast}$ as the value of $n$ where $P_0$ reaches $1.0$, we found that $n^{\ast}$ of PT is 322,800, while $n^{\ast}$ of \modelMultiQ~is only 600. In fact, for \modelMultiQ~the ground state can be found with only one gauge randomization for some instances, i.e., $P_0 > 0$ for $n=m$, with $m$ the number of its rounds.

\subsection*{Performance of minimizing the energy}
For larger systems, it is hard to directly compare the probability of finding the ground states, because even the best branch-and-bound solver could not calculate the ground states within acceptable time. A more practical choice of benchmarking various methods is to compare the energy of the ``ground state", denoted as $E_0$, predicted by each method. In particular, we are interested in the disorder averaged ``ground-state" energy per spin, denoted as $e_0$, which is computed as $E_0/N$ averaged over many instances. In Fig.~\ref{fig:convergence-comp}, we demonstrate $e_0$ as a function of the number of initial configurations $n$, we found that \modelMultiQ~maintains the lowest $e_0$ for almost all the dimensions and sizes. In some cases (e.g., $d=2$, $L=25$), \modelMultiQ~with only one gauge randomization (counts as $m=6$ initial configurations) could beat Greedy, SA and \modelMetaQ~with $2\times10^4$ initial configurations, and PT with 306 epochs (counts as $306\times20=6,120$ initial configurations). This fully verifies the effectiveness of \modelMultiQ~in minimizing the energy of EA spin glass models.

We notice a clear performance gap between \modelMetaQ~and \modelMultiQ~in Fig.~\ref{fig:convergence-comp}. Since \modelMultiQ~just repeats multiple rounds of \modelMetaQ, this gap actually tells us that \modelMetaQ~would not fall into the local optimum after a round like Greedy does. This might be another gain of \modelMetaQ's long-term sight. 

We also notice that although with increasing $n$, all methods eventually get lower $e_0$, but obviously PT displays the largest decrease of $e_0$. PT does not perform extremely well at first, but it gradually reaches lower and lower $e_0$ and eventually outperforms Greedy, SA and \modelMetaQ. In the case of $d=4$ and $L=6$ (Fig.~\ref{fig:convergence-comp} {\bf i}), PT eventually reaches an $e_0$ that is even lower than that of \modelMultiQ. However, as shown in Fig.~\ref{fig:convergence-comp} that \modelPT, as a \model-plug-in enhanced PT, significantly outperforms PT for all the dimensions and sizes tested here, especially when the number of initial configurations is relatively small. This suggests that replacing the Metropolis-Hasting criterion with a single \model~descent at low temperature significantly improves the performance of PT. 

\subsection*{Efficiency}
Besides the effectiveness, \model~is also computationally efficient. During the application phase, we flip a small finite fraction (e.g., 1\%) of the highest-$Q$ spins at each adaptive step, rather than the one-by-one flip in the training phase. In our numerical experiments, we found this \emph{batch nodes selection} strategy~\cite{fan2020finding} reduces the running time significantly without sacrificing much accuracy. Both time complexity and empirical calculations suggest that \model~with \emph{batch nodes selection} displays better scalability than other methods. 

We should admit that \model~needs to be offline trained while other methods do not have to. We think it is reasonable to compare \model's efficiency without considering its training time, since \model~needs to be trained offline only once for each dimension, (which all converge rapidly), and could then be applied infinite number of times. Besides, \model's training time is often affordable. For example, it takes about 2.5 hours for \model~to finish training on the 3D Ising spin glass system (all the calculations were conducted on a 20-core computer server with 512GB memory and a 16GB Tesla V100 GPU). For some large systems, the running time of \model, even with the training time considered, is still faster than that of PT. For example, the well-trained 3D \modelMultiQ~takes an average 417 seconds (one gauge randomization, counts as $n=m=8$ initial configurations, the same calculation costs only 252 seconds with graphics processing unit (GPU)-accelerations) to calculate a random spin glass instance with $L=20$, whilst PT algorithm needs an average 3 hours (191 epochs, counts as $n=191\times20=3,820$ initial configurations) to achieve the same energy (Fig.~\ref{fig:convergence-comp} {\bf f}). 
Furthermore, multiple gauge randomizations of \modelMultiQ~can be calculated with trivial parallelization, since these runs all start from independent configurations. However, multiple epochs of PT can only be computed in a serial mode since these epochs are consecutive with each other.

Since the biggest computational cost of \model~comes from the matrix multiplications in \ENCmodel, the GPU-accelerations can be more easily applied on \model~than other methods, as the matrix multiplication itself is particularly suitable for paralleling. Still, for the sake of fairness, here we report \model’s CPU running times only, and do not deploy its GPU-accelerations in the application phase. We only utilized GPU to speed up the training phase. Hence, the efficiency of \model~presented here is rather conservative. 

\subsection*{Training using different coupling distributions}
The results presented so far are for Gaussian EA spin glass instances, of which the couplings $\{J_{ij}\}$ are sampled from the Gaussian distribution. We used Gaussian, Bimodal and Uniform distributions to generate training and test instances separately, and analyze their impacts on \model. We found that \model~performs the best when test instances are generated from the same distribution as the training instances, and \model~trained with instances from the Gaussian distribution generalizes better than those trained with instances from the other two distributions.

\subsection*{Application on the max-cut problem}
Since many NP-hard problems have Ising spin glass formulations~\cite{lucas2014ising}, an efficient algorithm for the Ising spin glass ground state problem paves the way to better solutions of these problems. We applied \model~to solve the max-cut problem, a canonical example of the mapping between Ising spin glasses and NP-hard problems~\cite{lucas2014ising}. We found that \model~consistently outperforms other competing max-cut solvers. We anticipate that \model~could benefit us in solving a wide range of NP-hard problems that have Ising spin glass formulations.

%% file: main_text_section/conclusion.tex
\section*{Discussion}
This work reports an effective and efficient deep reinforcement learning based algorithm, \model, that can directly calculate the ground states of EA spin glasses. Extensive numerical calculations demonstrate that \model~outperforms state-of-the-art algorithms in terms of both solution quality and running time. Through a pure data-driven way and with no domain-specific guidance, \model~smartly learns to sacrifice in the short term to obtain long-term gains. The learned \model~enables a much faster finding of true ground states than existing algorithms, and it maintains the lowest system energy for almost all dimensions and sizes. Besides, \model~could also serve as a general plug-in and improve Monte-Carlo based methods, e.g., parallel-tempering. We believe that \model~will facilitate us solve many mysteries behind the strange and complex behaviors of spin glasses. It will also  help us solve many other NP-hard problems with Ising spin glass formulations. Moreover, the gauge transformation adopted in \model~might open up a new promising avenue for reinforcement learning models to be trained in the finite-horizon MDP and applied in the infinite-horizon MDP, which would be extremely beneficial to its use in other NP-hard combinatorial optimization problems.

We openly admit that this work has some limitations. For example, compared to the state-of-the-art PT algorithm, \modelMultiQ~converges much faster as the number of initial configuration increases, this hinders \modelMultiQ~from achieving lower energy (for high dimensions and large sizes), as the convergence also means the reaching of local optimum. By contrast, PT converges more slowly, resulting in a lower energy for some high dimensions and large sizes. We think this is largely due to the art of PT's balance between high-temperature configuration space exploration and low-temperature energy descent. During each epoch, PT keeps exploring the whole configuration space by accepting all possible moves at high-temperature replicas, and searching for lower-energy configurations at low-temperature replicas. At the end of each epoch, PT exchanges the configurations of high-temperature replicas with the low-temperature ones, which manages to prevent the low-temperature replicas from getting stuck in some local minimums. The diversity of the initial spin configurations makes PT more likely to find out the ground state in the end. \modelPT~can be regarded as an improvement of \model~in this respect, which balance between high-temperature configuration space exploration and low-temperature \model~descent exploitation. We can see \modelPT~converges slower than both \modelMetaQ~and \modelMultiQ, and could consistently outperform PT. However, the gap between PT and \modelPT~gradually diminishes as the number of configurations increases. 

In the future, we need to explore more effective ways to combine PT and \model, so as to better approach the ground states with large amounts of initial configurations. Besides, advances in deep graph representations may enable us design better encoder and further improve \model's performances to find the ground states of Ising spin glasses. Our current framework is just the beginning of such a promising adventure.

%% file: main_text_section/figures.tex
%%%%%%%%%%%%%%% Figure 1 %%%%%%%%%%%%%%%%%
\clearpage
\begin{figure}
\centering  
\label{fig:examples}
\subfigure{ 
\label{fig:b}     
\includegraphics[width=0.23\columnwidth]{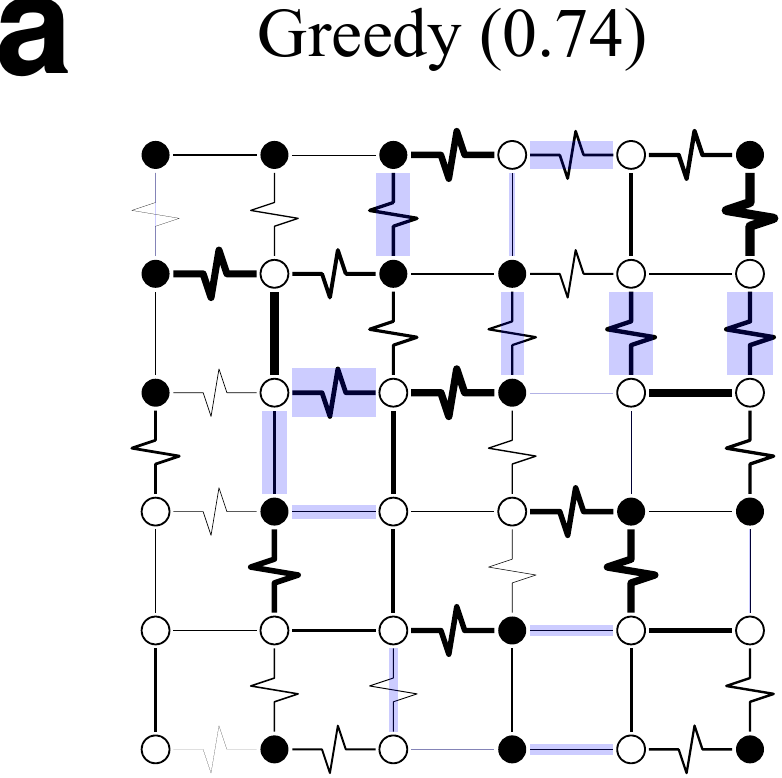}     
}    
\subfigure{ 
\label{fig:c}     
\includegraphics[width=0.23\columnwidth]{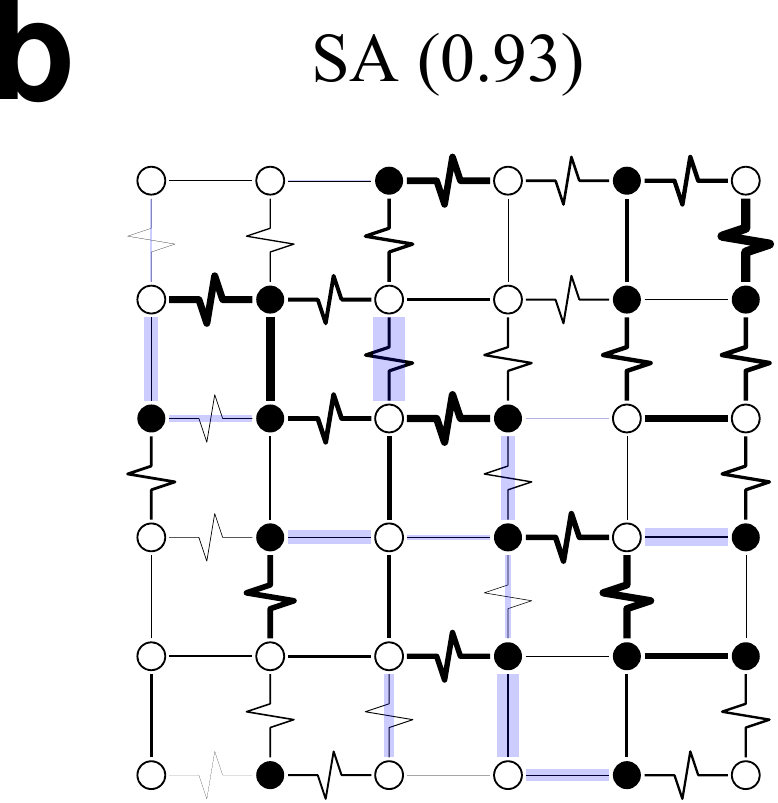}     
} 
\subfigure{ 
\label{fig:d}     
\includegraphics[width=0.23\columnwidth]{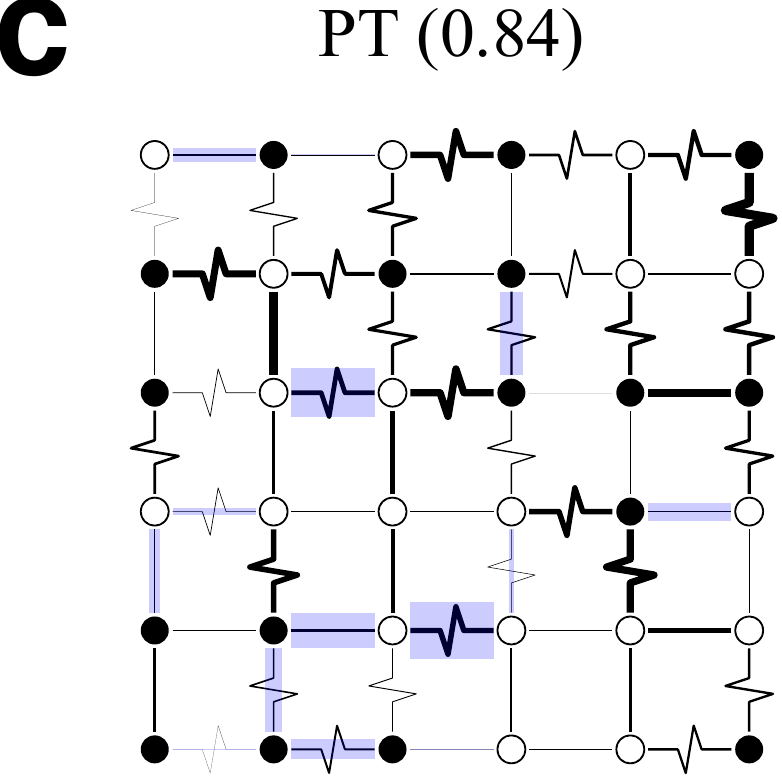}     
}
\subfigure{ 
\label{fig:e}     
\includegraphics[width=0.23\columnwidth]{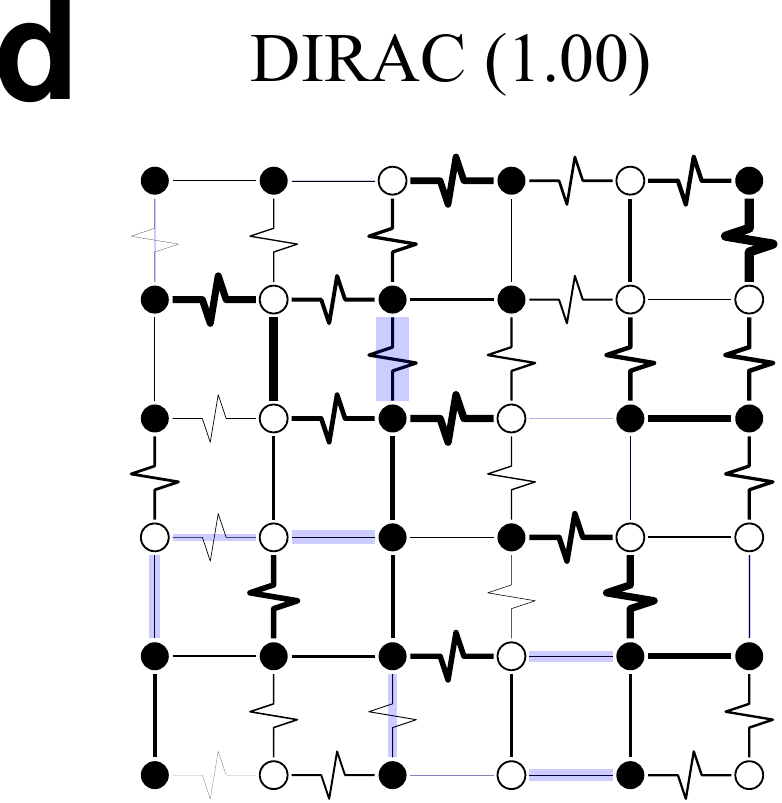}     
}
\caption[Case study comparison]{{\noindent \bf Figure 1 \hspace{2mm} Case study comparison.} We applied four algorithms: (\textbf{a}) Greedy, (\textbf{b}) Simulated annealing (SA), (\textbf{c}) Parallel tempering (PT), and ({\bf d}) \model~(more precisely, \modelMultiQ) in finding the ground state of a randomly generated $6 \times 6$ Edwards–Anderson (EA) spin glass instance with fixed boundary conditions and couplings $J_{ij}$ sampled from the Gaussian distribution $\mathcal{N}(0,1)$. The ferromagnetic bonds ($J_{ij}>0$) are shown with straight lines, while the anti-ferromagnetic bonds ($J_{ij}<0$) are shown with zigzag lines. The width of the lines are proportional to $|J_{ij}|$. Nodes are filled/hollow if the spin values $\sigma_i=+1/-1$, respectively. If the energy of a bond, $-J_{ij}s_is_j$, is positive, namely not satisfied, we draw a light blue shaded rectangle around the bond, with width proportional to $|J_{ij}|$. This way smaller total shaded area of the image corresponds to lower system energies. We also show the approximate ratio ($\frac{\text{prediction}}{\text{ground truth}}$, where the numerator is the energy of the predicted ground state computed by each method and the denominator is the exact ground state energy computed by the Gurobi solver~\cite{gurobi}) in brackets. Note that in this small case \model~can actually achieve the exact ground state.}  
\label{fig:case_study}     
\end{figure}

%%%%%%%%%%%%%%% Figure 2 %%%%%%%%%%%%%%%%%
\clearpage
\begin{figure}
\centering
\includegraphics[width=1.0\textwidth]{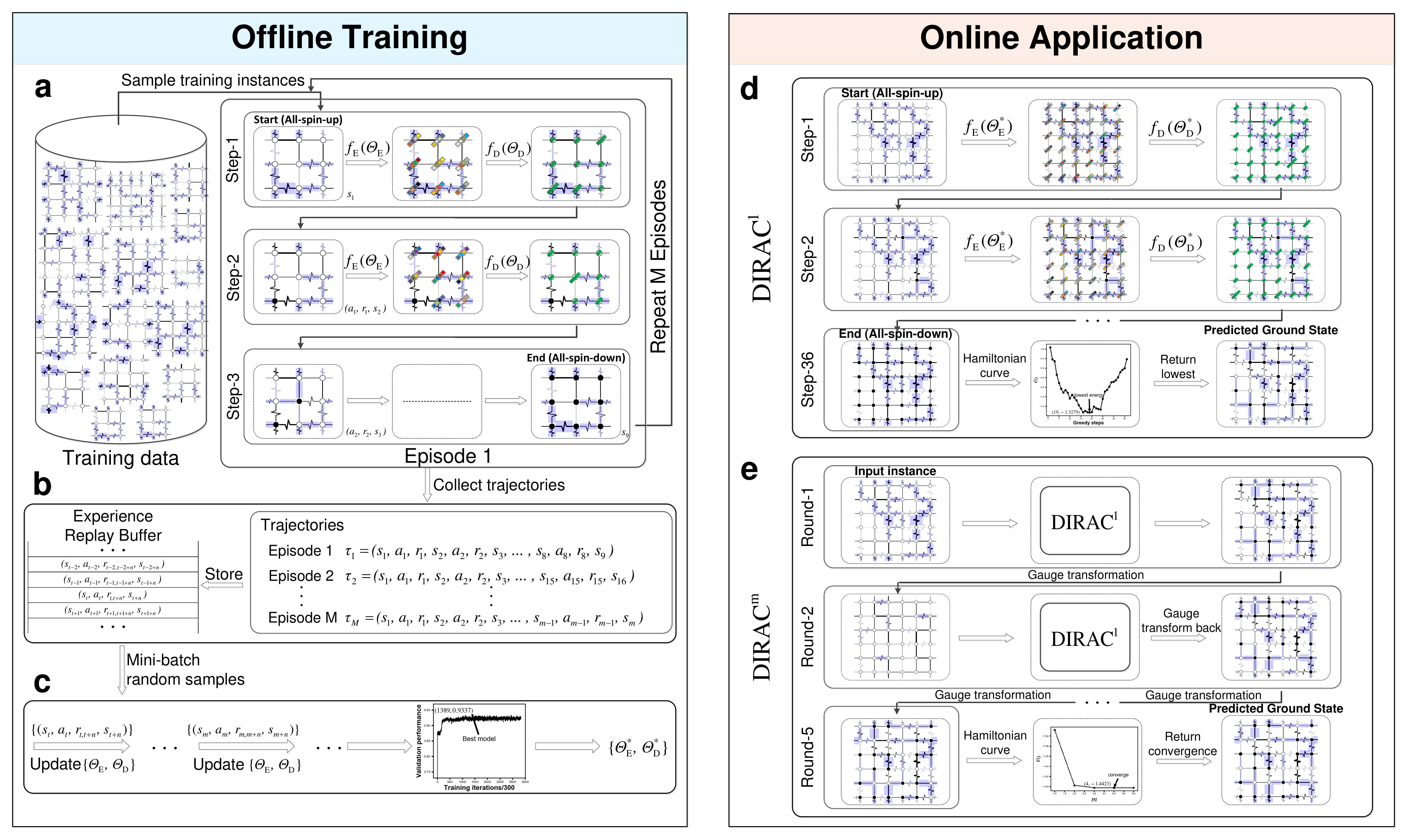}
 \caption[Overview of \model]{{\noindent \bf Figure 2 \hspace{2mm} Overview of \model.} The \model~framework is composed of two phases: the offline training phase and the online application phase. ({\bf Left}) During the training phase, we first generate random EA spin glass instances with couplings sampled from a certain distribution (e.g., Gaussian distribution), and then train for many episodes on these instances. {\bf a}, An \emph{episode} is the entire ground states finding process which starts from the all-spin-up configuration and ends at the all spin-down configuration with each spin flipped only once, and during which a trajectory of \emph{state}, \emph{action} and \emph{reward} is generated ({\bf b}). Here, \emph{state} is the observed instance, \emph{action} is to flip a spin, and \emph{reward} is the energy change after taking the action. To determine the right action to take, \model~first adopts an encoder \ENCmodel~to represent each node as a low-dimensional embedding vector (shown as color bars), and then leverage these representations and decode a $Q$ score (shown as green bars, with heights proportional to its values) for each node that predicts its long-term rewards. To balance between exploration and exploitation, we adopt the $\epsilon$-greedy strategy, which }
\label{fig:framework}
\end{figure}
\clearpage
\noindent takes the highest-$Q$ node with probability (1-$\epsilon$) and random actions otherwise, with $\epsilon$ linearly annealing from 1.0 to 0.05 over 50,000 episodes. After an episode ends, we obtain a trajectory ($s_1, a_1, r_1,…,s_l$), and collect the $n$-step transitions, i.e., $(s_t, a_t, r_{t,t+n}, s_{t+n})$, where $r_{t,t+n}=\sum_{k=t}^{t+n}r_k$, into the experiences replay buffer $\mathcal{B}$, which is a queue that maintains a number of mot recent $n$-step transitions. {\bf b}, To update parameters, we randomly sample mini-batch transitions from the buffer and perform gradient descents over Eq.~(\ref{loss}). {\bf c}, Repeating the episodes and updates until the training process converges (measured by the validation performance), we obtain the optimized parameters $\{\Theta^{\ast}_{\rm E}, \Theta^{\ast}_{\rm D}\}$, which can be applied directly in the application phase. ({\bf Right}) During application, we have two strategies: \modelMetaQ~and \modelMultiQ. {\bf d}, \modelMetaQ~performs like training, which starts from the all-spin-up configuration for an input spin glass instance, and ends at the all-spin-down configuration. The differences lie in that \modelMetaQ~adopts the batch $Q^{\ast}$ greedy strategy, which flips a finite fraction (e.g., 1\%) of highest $Q^{\ast}$ spins each step, rather than the one-by-one $\epsilon$-greedy flip in training. The spin configuration of the lowest energy encountered during this process is returned as the predicted ground state. {\bf e}, \modelMultiQ~can be considered as a multi-round-\modelMetaQ, which repeats multiple rounds of \modelMetaQ~for a given instance until the system energy converges. For each round, \modelMultiQ~takes input the all-spin-up configuration that is gauge transformed from the returned configuration in last round, then repeats \modelMetaQ, the returned configuration should be gauge transformed back to be the predicted ground state of this round. Here the gauge transformation technique switches the spin glass system between any two configurations while keeping the system energy invariant~\cite{Ozeki95,PhysRevB.72.045137}. The spin configuration of the converged energy is returned as the predicted ground state of \modelMultiQ. Note that with gauge transformation, we could also perform multiple gauge randomizations for both \modelMetaQ~and \modelMultiQ, with each randomization starting from a different initial configuration, and return the lowest energy among these runs.

%%%%%%%%%%%%%%% Figure 2 %%%%%%%%%%%%%%%%%
\clearpage
\begin{figure}
\centering
\includegraphics[width=0.75\textwidth]{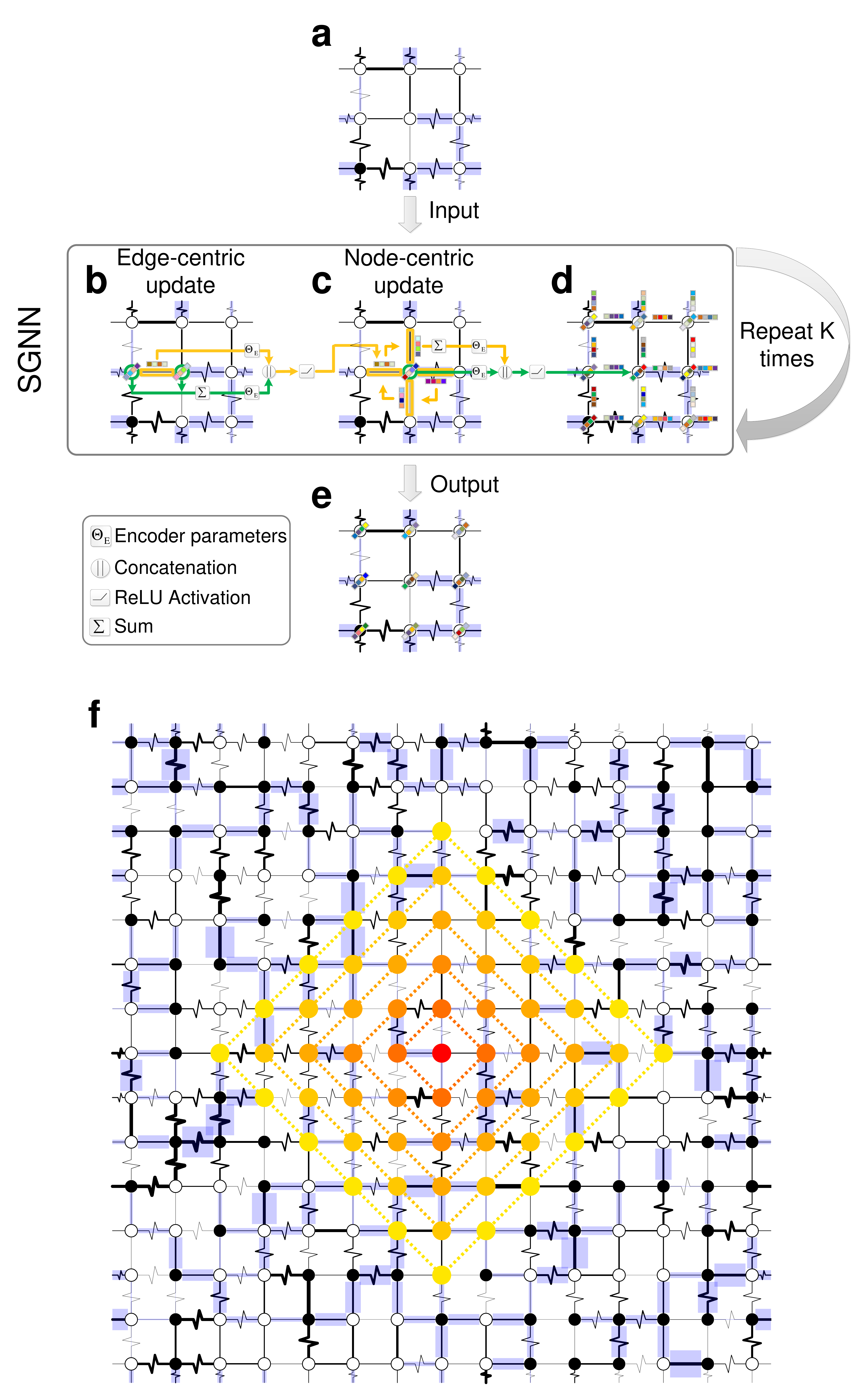}
 \caption[Spin glass Neural Network (\ENCmodel) encoder]{{\noindent \bf Figure 3 \hspace{2mm} \ENCmodel~encoder.} \ENCmodel~encodes ({\bf a}) the input spin glass instance (with any arbitrary spin configuration) into a ({\bf e}) low-dimensional space, where each spin is associated with an}\label{fig:SGNN}
\end{figure}
\clearpage
\noindent {embedding vector (shown as color bars). ({\bf b}) \ENCmodel~first updates edge embedding vectors based on the edge itself and its adjacent nodes, and then ({\bf c}) updates node embedding vectors based on the node itself and its adjacent edges. Note that the node-centric updates take place only when the edge-centric updates finish for all edges at each layer. Both updates are followed by a non-linear transformation operator (e.g., $\mathrm{ReLU}$) with learnable parameters. The edge features are initialized by edge weights, and the node features are initialized by its coordinates in the hypercubic coordinate system. {\bf d}, Each node or edge updates its embedding vector in one layer. Repeating several layers, we obtain an embedding vector for each node ({\bf e}) that reflects its informative features. {\bf f}, Each layer of updates increases the long-range couplings with one more hop's neighbors (dashed lines) for a given spin. For example, for the central spin (colored in dark red), its final embedding vector after $K=5$ layers captures both its position and its long-range couplings with neighbors within $K=5$ hops.}

%%%%%%%%%%%%%%% Figure 3 %%%%%%%%%%%%%%%%%
\clearpage
\begin{figure}
\centering
\includegraphics[width=1.0\textwidth]{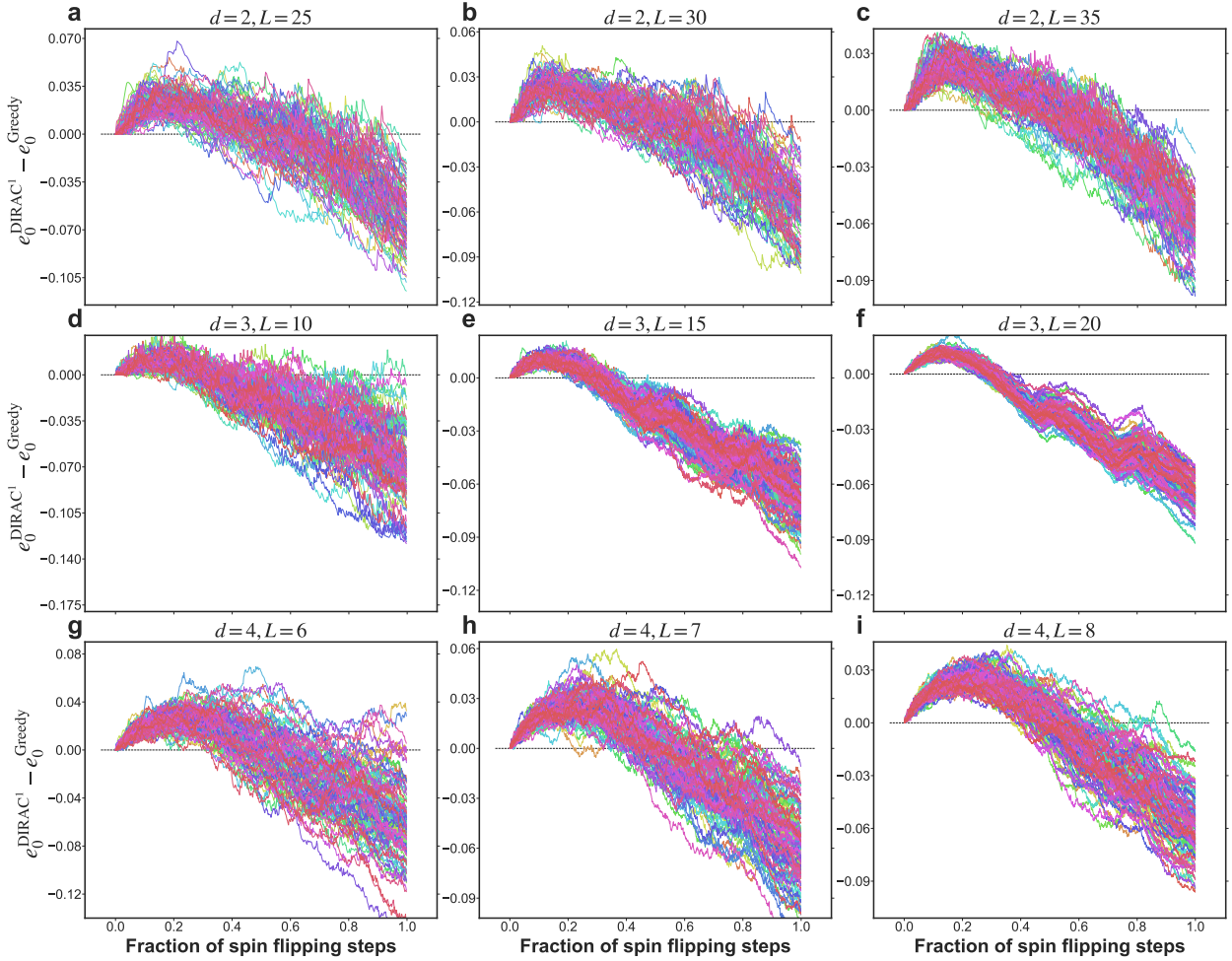}
 \caption[A long-sighted greediness]{{\noindent \bf Figure 4 \hspace{2mm} A long-sighted greediness.} We compare the energy difference between \modelMetaQ~and the Greedy algorithm at each step. Note that the greedy steps of two methods may not be exactly the same, we make up the length of the shorter sequence with its last value, so as the two sequences can be compared step by step. Note that to compare with Greedy step by step more precisely, here the results of \modelMetaQ~are achieved by flipping only one spin each step. (\textbf{a}-\textbf{i}) illustrate the energy gaps for different dimensions and different sizes. For each size, we compare on 100 randomly generated EA spin glass instances (with couplings sampled from Gaussian distribution $\mathcal{N}(0,1)$), which are represented by 100 curves of different colors. It can be clearly seen that \modelMetaQ~always goes through a high-energy state temporarily in the early stage of the greedy process, so as to reach a much lower energy state in the long run.}
\label{fig:energy_gap}
\end{figure}

%%%%%%%%%%%%%%% Figure 3 %%%%%%%%%%%%%%%%%
\clearpage
\begin{figure}
\centering
\includegraphics[width=1.0\textwidth]{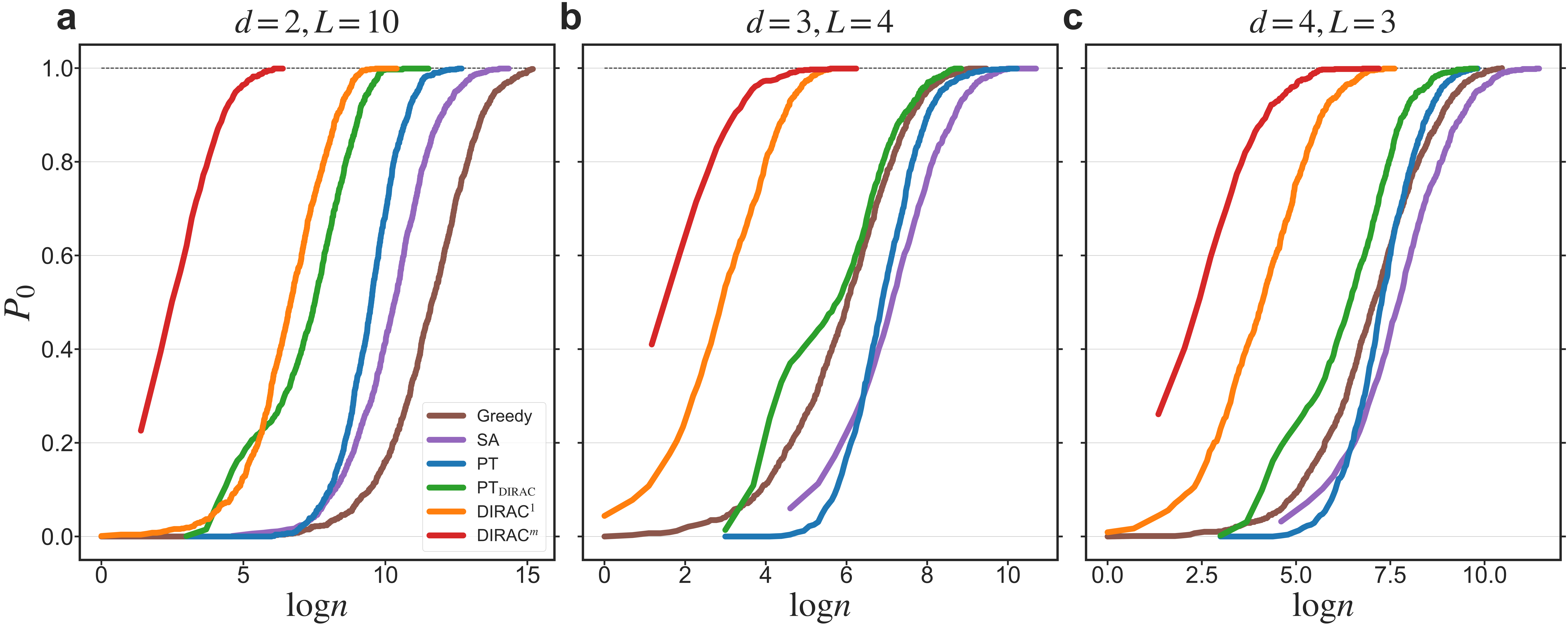}
 \caption[Performance of finding the ground state]{{\noindent \bf Figure 5 \hspace{2mm} Performance of finding the ground state.} To compare the ability of finding the exact ground states, we evaluate different methods on small spin glass systems for which the true ground state can be computed by the branch-and-bound-based Gurobi solver~\cite{gurobi}. The quantity we choose to compare is the mean probability of finding the ground state $P_0$, which is calculated as the fraction of 1000 random instances for which the ground state is found. Note that it is not necessary to consider many runs for one instance, the error associated $P_0$ becomes small if many instances are calculated, and only one run is enough for each one of them~\cite{roma2009ground}. We compare $P_0$ as a function of the number of different initial configurations $n$ (log scale is used here for visualization purpose). For PT, \modelPT~and \modelMultiQ, $n=tm$, where $t$ is the number of PT (or \modelPT) epochs or \modelMultiQ~gauge randomizations, $m$ is the number of PT (or \modelPT) replicas (set as 20) or number of \modelMultiQ~rounds. For \modelMetaQ, $n$ just refers to the number of independent gauge randomizations. Note that PT curves in ({\bf a}) and ({\bf b}) show slight shifts to the right compared to Fig.~6 and Fig.~11 in Ref.~\cite{roma2009ground}, it is because the $t$ in x-axis in Ref.~\cite{roma2009ground} is actually the $t/2.3$ in our case (we implement the B variant of PT algorithm in \cite{roma2009ground}).}
\label{fig:gs_compare}
\end{figure}

%%%%%%%%%%%%%%% Figure 4 %%%%%%%%%%%%%%%%%
\clearpage
\begin{figure}
\centering
\includegraphics[width=1.0\textwidth]{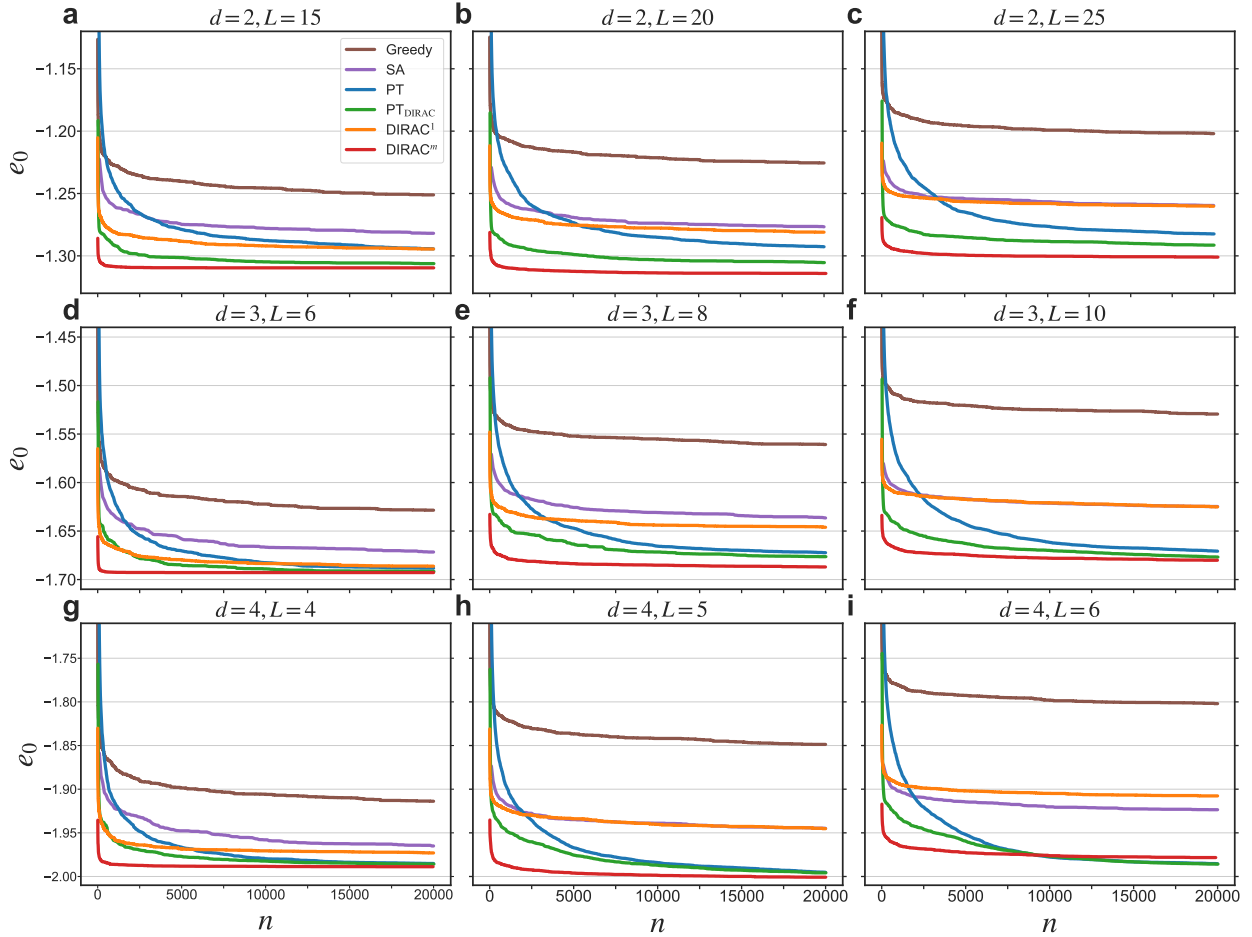}
 \caption[Performance of minimizing the energy]{{\noindent \bf Figure 6 \hspace{2mm} Performance of minimizing the energy.} We compare the disorder averaged ``ground-state" energy per spin (predicted by each method), denote as $e_0$, as a function of the number of initial configurations $n$, to benchmark various methods on large systems. We compare on three dimensions ($d=2,3,4$), and for each dimension we consider three different sizes ({\bf a-i}). For each case, we run $n=2\times10^4$ initial configurations for each method to reach its approximate convergence. At a given $n$, we choose the lowest energy among all runs for each instance, and average the 50 independent instances as the final result. Note that some literature reported a lower PT energy than what we show here, e.g., \cite{roma2009ground, ComparingMethods} calculates an average energy $-1.6981$ for the system with $d=3$ and $L=10$, while we calculate only $-1.6707$ ({\bf f}). We claim this is because they run}
\label{fig:convergence-comp}
\end{figure}
\clearpage
\noindent with up to $3.2\times10^7$ initial configurations, while we only compute up to $n=2\times10^4$ ones due to high computational costs.

%% file: main_text.bbl
\begin{thebibliography}{10}
\expandafter\ifx\csname url\endcsname\relax
  \def\url#1{\texttt{#1}}\fi
\expandafter\ifx\csname urlprefix\endcsname\relax\def\urlprefix{URL }\fi
\providecommand{\bibinfo}[2]{#2}
\providecommand{\eprint}[2][]{\url{#2}}

\bibitem{binder1986spin}
\bibinfo{author}{Binder, K.} \& \bibinfo{author}{Young, A.~P.}
\newblock \bibinfo{title}{Spin glasses: Experimental facts, theoretical
  concepts, and open questions}.
\newblock \emph{\bibinfo{journal}{Reviews of Modern Physics}}
  \textbf{\bibinfo{volume}{58}}, \bibinfo{pages}{801} (\bibinfo{year}{1986}).

\bibitem{mezard1987spin}
\bibinfo{author}{M{\'e}zard, M.}, \bibinfo{author}{Parisi, G.} \&
  \bibinfo{author}{Virasoro, M.}
\newblock \emph{\bibinfo{title}{Spin glass theory and beyond: An Introduction
  to the Replica Method and Its Applications}}, vol.~\bibinfo{volume}{9}
  (\bibinfo{publisher}{World Scientific Publishing Company},
  \bibinfo{year}{1987}).

\bibitem{hartmann2002optimization}
\bibinfo{author}{Hartmann, A.~K.} \& \bibinfo{author}{Rieger, H.}
\newblock \emph{\bibinfo{title}{Optimization algorithms in physics}},
  vol.~\bibinfo{volume}{2} (\bibinfo{publisher}{Wiley Online Library},
  \bibinfo{year}{2002}).

\bibitem{sherrington1975solvable}
\bibinfo{author}{Sherrington, D.} \& \bibinfo{author}{Kirkpatrick, S.}
\newblock \bibinfo{title}{Solvable model of a spin-glass}.
\newblock \emph{\bibinfo{journal}{Physical Review Letters}}
  \textbf{\bibinfo{volume}{35}}, \bibinfo{pages}{1792} (\bibinfo{year}{1975}).

\bibitem{Chayes86amean}
\bibinfo{author}{Chayes, J.~T.}, \bibinfo{author}{I, L.~C.},
  \bibinfo{author}{Sethna, J.~P.} \& \bibinfo{author}{Thouless, D.~J.}
\newblock \bibinfo{title}{A mean field spin glass with short range
  interactions}.
\newblock \emph{\bibinfo{journal}{Comm. Math. Phys}}  (\bibinfo{year}{1986}).

\bibitem{ceccarelli2011ferromagnetic}
\bibinfo{author}{Ceccarelli, G.}, \bibinfo{author}{Pelissetto, A.} \&
  \bibinfo{author}{Vicari, E.}
\newblock \bibinfo{title}{Ferromagnetic-glassy transitions in three-dimensional
  ising spin glasses}.
\newblock \emph{\bibinfo{journal}{Physical Review B}}
  \textbf{\bibinfo{volume}{84}}, \bibinfo{pages}{134202}
  (\bibinfo{year}{2011}).

\bibitem{cugliandolo1995weak}
\bibinfo{author}{Cugliandolo, L.~F.} \& \bibinfo{author}{Kurchan, J.}
\newblock \bibinfo{title}{Weak ergodicity breaking in mean-field spin-glass
  models}.
\newblock \emph{\bibinfo{journal}{Philosophical Magazine B}}
  \textbf{\bibinfo{volume}{71}}, \bibinfo{pages}{501--514}
  (\bibinfo{year}{1995}).

\bibitem{carter2002aspect}
\bibinfo{author}{Carter, A.}, \bibinfo{author}{Bray, A.} \&
  \bibinfo{author}{Moore, M.}
\newblock \bibinfo{title}{Aspect-ratio scaling and the stiffness exponent
  $\theta$ for ising spin glasses}.
\newblock \emph{\bibinfo{journal}{Physical Review Letters}}
  \textbf{\bibinfo{volume}{88}}, \bibinfo{pages}{077201}
  (\bibinfo{year}{2002}).

\bibitem{hartmann1999scaling}
\bibinfo{author}{Hartmann, A.~K.}
\newblock \bibinfo{title}{Scaling of stiffness energy for
  three-dimensional$\pm$j ising spin glasses}.
\newblock \emph{\bibinfo{journal}{Physical Review E}}
  \textbf{\bibinfo{volume}{59}}, \bibinfo{pages}{84} (\bibinfo{year}{1999}).

\bibitem{barahona1982computational}
\bibinfo{author}{Barahona, F.}
\newblock \bibinfo{title}{On the computational complexity of ising spin glass
  models}.
\newblock \emph{\bibinfo{journal}{Journal of Physics A: Mathematical and
  General}} \textbf{\bibinfo{volume}{15}}, \bibinfo{pages}{3241}
  (\bibinfo{year}{1982}).

\bibitem{lucas2014ising}
\bibinfo{author}{Lucas, A.}
\newblock \bibinfo{title}{Ising formulations of many np problems}.
\newblock \emph{\bibinfo{journal}{Frontiers in Physics}}
  \textbf{\bibinfo{volume}{2}}, \bibinfo{pages}{5} (\bibinfo{year}{2014}).

\bibitem{Hopfield}
\bibinfo{author}{Hopfield, J.~J.}
\newblock \bibinfo{title}{Neural networks and physical systems with emergent
  collective computational abilities}.
\newblock \emph{\bibinfo{journal}{Proceedings of the National Academy of
  Sciences (USA)}} \textbf{\bibinfo{volume}{79}}, \bibinfo{pages}{2554}
  (\bibinfo{year}{1982}).

\bibitem{Little}
\bibinfo{author}{Little, W.~A.}
\newblock \bibinfo{title}{The existence of persistent states in the brain}.
\newblock \emph{\bibinfo{journal}{Math. Biosci.}}
  \textbf{\bibinfo{volume}{19}}, \bibinfo{pages}{101} (\bibinfo{year}{1974}).

\bibitem{Amit}
\bibinfo{author}{Amit, D.~J.}, \bibinfo{author}{Gutfreund, H.} \&
  \bibinfo{author}{Sompolinsky, H.}
\newblock \bibinfo{title}{Spin-glass models of neural networks}.
\newblock \emph{\bibinfo{journal}{Physical Review A}}
  \textbf{\bibinfo{volume}{32}}, \bibinfo{pages}{1007} (\bibinfo{year}{1985}).

\bibitem{Sompolinsky}
\bibinfo{author}{Sompolinsky, H.}
\newblock \bibinfo{title}{Statistical mechanics of neural networks}.
\newblock \emph{\bibinfo{journal}{Physics Today}}
  \textbf{\bibinfo{volume}{40}}, \bibinfo{pages}{70} (\bibinfo{year}{1988}).

\bibitem{Fortunato}
\bibinfo{author}{Fortunato, S.}
\newblock \bibinfo{title}{Community detection in graphs}.
\newblock \emph{\bibinfo{journal}{Physics Reports}}
  \textbf{\bibinfo{volume}{486}}, \bibinfo{pages}{75} (\bibinfo{year}{2010}).

\bibitem{RB}
\bibinfo{author}{Reichardt, J.} \& \bibinfo{author}{Bornholdt, S.}
\newblock \bibinfo{title}{Statistical mechanics of community detection}.
\newblock \emph{\bibinfo{journal}{Physical Review E}}
  \textbf{\bibinfo{volume}{74}}, \bibinfo{pages}{016110}
  (\bibinfo{year}{2006}).

\bibitem{BP1}
\bibinfo{author}{M{\'e}zard, M.}, \bibinfo{author}{Parisi, G.} \&
  \bibinfo{author}{Zecchina, R.}
\newblock \bibinfo{title}{Analytic and algorithmic solution of random
  satisfiability problems}.
\newblock \emph{\bibinfo{journal}{Science}} \textbf{\bibinfo{volume}{297}},
  \bibinfo{pages}{812--815} (\bibinfo{year}{2002}).

\bibitem{de1995exact}
\bibinfo{author}{De~Simone, C.} \emph{et~al.}
\newblock \bibinfo{title}{Exact ground states of ising spin glasses: New
  experimental results with a branch-and-cut algorithm}.
\newblock \emph{\bibinfo{journal}{Journal of Statistical Physics}}
  \textbf{\bibinfo{volume}{80}}, \bibinfo{pages}{487--496}
  (\bibinfo{year}{1995}).

\bibitem{hartmann2011ground}
\bibinfo{author}{Hartmann, A.~K.}
\newblock \bibinfo{title}{Ground states of two-dimensional ising spin glasses:
  fast algorithms, recent developments and a ferromagnet-spin glass mixture}.
\newblock \emph{\bibinfo{journal}{Journal of Statistical Physics}}
  \textbf{\bibinfo{volume}{144}}, \bibinfo{pages}{519} (\bibinfo{year}{2011}).

\bibitem{weigel2018}
\bibinfo{author}{Khoshbakht, H.} \& \bibinfo{author}{Weigel, M.}
\newblock \bibinfo{title}{Domain-wall excitations in the two-dimensional ising
  spin glass}.
\newblock \emph{\bibinfo{journal}{Physical Review B}}
  \textbf{\bibinfo{volume}{97}}, \bibinfo{pages}{064410}
  (\bibinfo{year}{2018}).

\bibitem{kirkpatrick1983optimization}
\bibinfo{author}{Kirkpatrick, S.}, \bibinfo{author}{Gelatt, C.~D.} \&
  \bibinfo{author}{Vecchi, M.~P.}
\newblock \bibinfo{title}{Optimization by simulated annealing}.
\newblock \emph{\bibinfo{journal}{Science}} \textbf{\bibinfo{volume}{220}},
  \bibinfo{pages}{671--680} (\bibinfo{year}{1983}).

\bibitem{gubernatis2003monte}
\bibinfo{author}{Gubernatis, J.~E.}
\newblock \bibinfo{title}{The monte carlo method in the physical sciences:
  celebrating the 50th anniversary of the metropolis algorithm}.
\newblock \emph{\bibinfo{journal}{The Monte Carlo Method in the Physical
  Sciences}} \textbf{\bibinfo{volume}{690}} (\bibinfo{year}{2003}).

\bibitem{swendsen1986replica}
\bibinfo{author}{Swendsen, R.~H.} \& \bibinfo{author}{Wang, J.-S.}
\newblock \bibinfo{title}{Replica monte carlo simulation of spin-glasses}.
\newblock \emph{\bibinfo{journal}{Physical Review Letters}}
  \textbf{\bibinfo{volume}{57}}, \bibinfo{pages}{2607} (\bibinfo{year}{1986}).

\bibitem{geyer1991computing}
\bibinfo{author}{Geyer, C.~J.} \emph{et~al.}
\newblock \bibinfo{title}{Computing science and statistics: Proceedings of the
  23rd symposium on the interface}.
\newblock \emph{\bibinfo{journal}{American Statistical Association, New York}}
  \textbf{\bibinfo{volume}{156}} (\bibinfo{year}{1991}).

\bibitem{hukushima1996exchange}
\bibinfo{author}{Hukushima, K.} \& \bibinfo{author}{Nemoto, K.}
\newblock \bibinfo{title}{Exchange monte carlo method and application to spin
  glass simulations}.
\newblock \emph{\bibinfo{journal}{Journal of the Physical Society of Japan}}
  \textbf{\bibinfo{volume}{65}}, \bibinfo{pages}{1604--1608}
  (\bibinfo{year}{1996}).

\bibitem{earl2005parallel}
\bibinfo{author}{Earl, D.~J.} \& \bibinfo{author}{Deem, M.~W.}
\newblock \bibinfo{title}{Parallel tempering: Theory, applications, and new
  perspectives}.
\newblock \emph{\bibinfo{journal}{Physical Chemistry Chemical Physics}}
  \textbf{\bibinfo{volume}{7}}, \bibinfo{pages}{3910--3916}
  (\bibinfo{year}{2005}).

\bibitem{mnih2015human}
\bibinfo{author}{Mnih, V.} \emph{et~al.}
\newblock \bibinfo{title}{Human-level control through deep reinforcement
  learning}.
\newblock \emph{\bibinfo{journal}{Nature}} \textbf{\bibinfo{volume}{518}},
  \bibinfo{pages}{529--533} (\bibinfo{year}{2015}).

\bibitem{li2018combinatorial}
\bibinfo{author}{Li, Z.}, \bibinfo{author}{Chen, Q.} \&
  \bibinfo{author}{Koltun, V.}
\newblock \bibinfo{title}{Combinatorial optimization with graph convolutional
  networks and guided tree search}.
\newblock In \emph{\bibinfo{booktitle}{Advances in Neural Information
  Processing Systems}}, \bibinfo{pages}{539--548} (\bibinfo{year}{2018}).

\bibitem{fan2020finding}
\bibinfo{author}{Fan, C.}, \bibinfo{author}{Zeng, L.}, \bibinfo{author}{Sun,
  Y.} \& \bibinfo{author}{Liu, Y.-Y.}
\newblock \bibinfo{title}{Finding key players in complex networks through deep
  reinforcement learning}.
\newblock \emph{\bibinfo{journal}{Nature Machine Intelligence}}
  \textbf{\bibinfo{volume}{2}}, \bibinfo{pages}{317--324}
  (\bibinfo{year}{2020}).

\bibitem{bello2016neural}
\bibinfo{author}{Bello, I.}, \bibinfo{author}{Pham, H.}, \bibinfo{author}{Le,
  Q.~V.}, \bibinfo{author}{Norouzi, M.} \& \bibinfo{author}{Bengio, S.}
\newblock \bibinfo{title}{Neural combinatorial optimization with reinforcement
  learning}.
\newblock \emph{\bibinfo{journal}{arXiv preprint arXiv:1611.09940}}
  (\bibinfo{year}{2016}).

\bibitem{nazari2018reinforcement}
\bibinfo{author}{Nazari, M.}, \bibinfo{author}{Oroojlooy, A.},
  \bibinfo{author}{Snyder, L.} \& \bibinfo{author}{Tak{\'a}c, M.}
\newblock \bibinfo{title}{Reinforcement learning for solving the vehicle
  routing problem}.
\newblock In \emph{\bibinfo{booktitle}{Advances in Neural Information
  Processing Systems}}, \bibinfo{pages}{9839--9849} (\bibinfo{year}{2018}).

\bibitem{udrescu2020ai}
\bibinfo{author}{{Udrescu}, S.-M.} \& \bibinfo{author}{{Tegmark}, M.}
\newblock \bibinfo{title}{Ai feynman: A physics-inspired method for symbolic
  regression}.
\newblock \emph{\bibinfo{journal}{Science Advances}}
  \textbf{\bibinfo{volume}{6}} (\bibinfo{year}{2020}).

\bibitem{PhysRevLett.122.080602}
\bibinfo{author}{Wu, D.}, \bibinfo{author}{Wang, L.} \& \bibinfo{author}{Zhang,
  P.}
\newblock \bibinfo{title}{Solving statistical mechanics using variational
  autoregressive networks}.
\newblock \emph{\bibinfo{journal}{Phys. Rev. Lett.}}
  \textbf{\bibinfo{volume}{122}}, \bibinfo{pages}{080602}
  (\bibinfo{year}{2019}).

\bibitem{mills2020finding}
\bibinfo{author}{Mills, K.}, \bibinfo{author}{Ronagh, P.} \&
  \bibinfo{author}{Tamblyn, I.}
\newblock \bibinfo{title}{Finding the ground state of spin hamiltonians with
  reinforcement learning}.
\newblock \emph{\bibinfo{journal}{Nature Machine Intelligence}}
  \textbf{\bibinfo{volume}{2}}, \bibinfo{pages}{509--517}
  (\bibinfo{year}{2020}).

\bibitem{silver2016mastering}
\bibinfo{author}{Silver, D.} \emph{et~al.}
\newblock \bibinfo{title}{Mastering the game of go with deep neural networks
  and tree search}.
\newblock \emph{\bibinfo{journal}{Nature}} \textbf{\bibinfo{volume}{529}},
  \bibinfo{pages}{484--489} (\bibinfo{year}{2016}).

\bibitem{khalil2017learning}
\bibinfo{author}{Khalil, E.}, \bibinfo{author}{Dai, H.},
  \bibinfo{author}{Zhang, Y.}, \bibinfo{author}{Dilkina, B.} \&
  \bibinfo{author}{Song, L.}
\newblock \bibinfo{title}{Learning combinatorial optimization algorithms over
  graphs}.
\newblock In \emph{\bibinfo{booktitle}{Advances in Neural Information
  Processing Systems}}, \bibinfo{pages}{6348--6358} (\bibinfo{year}{2017}).

\bibitem{mazyavkina2021reinforcement}
\bibinfo{author}{Mazyavkina, N.}, \bibinfo{author}{Sviridov, S.},
  \bibinfo{author}{Ivanov, S.} \& \bibinfo{author}{Burnaev, E.}
\newblock \bibinfo{title}{Reinforcement learning for combinatorial
  optimization: A survey}.
\newblock \emph{\bibinfo{journal}{Computers \& Operations Research}}
  \bibinfo{pages}{105400} (\bibinfo{year}{2021}).

\bibitem{kipf2016semi}
\bibinfo{author}{Kipf, T.~N.} \& \bibinfo{author}{Welling, M.}
\newblock \bibinfo{title}{Semi-supervised classification with graph
  convolutional networks}.
\newblock In \emph{\bibinfo{booktitle}{International Conference on Learning
  Representations}} (\bibinfo{year}{2017}).

\bibitem{hanjunnips2017}
\bibinfo{author}{Khalil, E.}, \bibinfo{author}{Dai, H.},
  \bibinfo{author}{Zhang, Y.}, \bibinfo{author}{Dilkina, B.} \&
  \bibinfo{author}{Song, L.}
\newblock \bibinfo{title}{Learning combinatorial optimization algorithms over
  graphs}.
\newblock In \emph{\bibinfo{booktitle}{Advances in Neural Information
  Processing Systems}}, \bibinfo{pages}{6348--6358} (\bibinfo{year}{2017}).

\bibitem{hamilton2017inductive}
\bibinfo{author}{Hamilton, W.}, \bibinfo{author}{Ying, Z.} \&
  \bibinfo{author}{Leskovec, J.}
\newblock \bibinfo{title}{Inductive representation learning on large graphs}.
\newblock In \emph{\bibinfo{booktitle}{Advances in Neural Information
  Processing Systems}}, \bibinfo{pages}{1024--1034} (\bibinfo{year}{2017}).

\bibitem{velickovic2017graph}
\bibinfo{author}{Velickovic, P.} \emph{et~al.}
\newblock \bibinfo{title}{Graph attention networks}.
\newblock In \emph{\bibinfo{booktitle}{International Conference on Learning
  Representations}} (\bibinfo{year}{2018}).

\bibitem{gilmer2017neural}
\bibinfo{author}{Gilmer, J.}, \bibinfo{author}{Schoenholz, S.~S.},
  \bibinfo{author}{Riley, P.~F.}, \bibinfo{author}{Vinyals, O.} \&
  \bibinfo{author}{Dahl, G.~E.}
\newblock \bibinfo{title}{Neural message passing for quantum chemistry}.
\newblock In \emph{\bibinfo{booktitle}{International Conference on Machine
  Learning}}, \bibinfo{pages}{1263--1272} (\bibinfo{organization}{PMLR},
  \bibinfo{year}{2017}).

\bibitem{xu2018powerful}
\bibinfo{author}{Xu, K.}, \bibinfo{author}{Hu, W.}, \bibinfo{author}{Leskovec,
  J.} \& \bibinfo{author}{Jegelka, S.}
\newblock \bibinfo{title}{How powerful are graph neural networks?}
\newblock In \emph{\bibinfo{booktitle}{International Conference on Learning
  Representations}} (\bibinfo{year}{2018}).

\bibitem{WegnerGauge}
\bibinfo{author}{Wegner, F.~J.}
\newblock \bibinfo{title}{Duality in generalized ising models and phase
  transitions without local order parameters}.
\newblock \emph{\bibinfo{journal}{Journal of Mathematical Physics}}
  \textbf{\bibinfo{volume}{12}}, \bibinfo{pages}{2259--2272}
  (\bibinfo{year}{1971}).

\bibitem{Ozeki95}
\bibinfo{author}{Ozeki, Y.}
\newblock \bibinfo{title}{Gauge transformation for dynamical systems of ising
  spin glasses}.
\newblock \emph{\bibinfo{journal}{Journal of Physics A: Mathematical and
  General}} \textbf{\bibinfo{volume}{28}}, \bibinfo{pages}{3645}
  (\bibinfo{year}{1995}).

\bibitem{PhysRevB.72.045137}
\bibinfo{author}{Batista, C.~D.} \& \bibinfo{author}{Nussinov, Z.}
\newblock \bibinfo{title}{Generalized elitzur's theorem and dimensional
  reductions}.
\newblock \emph{\bibinfo{journal}{Physical Review B}}
  \textbf{\bibinfo{volume}{72}}, \bibinfo{pages}{045137}
  (\bibinfo{year}{2005}).

\bibitem{hamze2018near}
\bibinfo{author}{Hamze, F.} \emph{et~al.}
\newblock \bibinfo{title}{From near to eternity: spin-glass planting, tiling
  puzzles, and constraint-satisfaction problems}.
\newblock \emph{\bibinfo{journal}{Physical Review E}}
  \textbf{\bibinfo{volume}{97}}, \bibinfo{pages}{043303}
  (\bibinfo{year}{2018}).

\bibitem{gurobi}
\bibinfo{author}{Gurobi~Optimization, L.}
\newblock \bibinfo{title}{Gurobi optimizer reference manual}
  (\bibinfo{year}{2021}).

\bibitem{roma2009ground}
\bibinfo{author}{Rom{\'a}, F.}, \bibinfo{author}{Risau-Gusman, S.},
  \bibinfo{author}{Ramirez-Pastor, A.~J.}, \bibinfo{author}{Nieto, F.} \&
  \bibinfo{author}{Vogel, E.~E.}
\newblock \bibinfo{title}{The ground state energy of the edwards--anderson spin
  glass model with a parallel tempering monte carlo algorithm}.
\newblock \emph{\bibinfo{journal}{Physica A: statistical mechanics and its
  applications}} \textbf{\bibinfo{volume}{388}}, \bibinfo{pages}{2821--2838}
  (\bibinfo{year}{2009}).

\bibitem{ComparingMethods}
\bibinfo{author}{Wang, W.}, \bibinfo{author}{Machta, J.} \&
  \bibinfo{author}{Katzgraber, H.~G.}
\newblock \bibinfo{title}{Comparing monte carlo methods for finding ground
  states of ising spin glasses: Population annealing, simulated annealing, and
  parallel tempering}.
\newblock \emph{\bibinfo{journal}{Physical Review E}}
  \textbf{\bibinfo{volume}{92}}, \bibinfo{pages}{013303}
  (\bibinfo{year}{2015}).

\end{thebibliography}
